\documentclass[journal,compsoc, 10pt,a4paper]{IEEEtran}
\usepackage{lipsum}

\usepackage[utf8]{inputenc}
\usepackage[T1]{fontenc}
\usepackage{graphicx}
\usepackage{dcolumn}
\usepackage{grffile}
\usepackage{longtable}
\usepackage{subcaption}
\usepackage{multirow}
\usepackage{float}
\usepackage{wrapfig}
\usepackage{rotating}
\usepackage[framemethod=tikz]{mdframed}
\usepackage[normalem]{ulem}
\usepackage{amsmath}
\usepackage{textcomp}
\usepackage{marvosym}
\usepackage{wasysym}
\usepackage[british]{babel}
\usepackage{amssymb}
\usepackage{capt-of}
\usepackage{booktabs}
\usepackage{listings}
\usepackage{color}          
\usepackage{zi4} 
\usepackage[numbers]{natbib}
\setcitestyle{square, comma, numbers,sort&compress, super}
\usepackage{hyperref}
\usepackage{amsmath,amssymb,amsfonts}
\usepackage{algorithmic}
\usepackage{graphicx}
\usepackage[disable]{todonotes}
\usepackage{textcomp}
\usepackage{xcolor}
\usepackage{makecell}
\hypersetup{draft}
\newcommand{\Mina}[1]{\todo[inline]{MS: #1}}
\newcommand{\Pontus}[1]{\todo[inline]{PS: #1}}

\graphicspath{ {./figures/} }
\definecolor{bluekeywords}{rgb}{0, 0, 0}
\definecolor{greencomments}{rgb}{0, 0.5, 0}
\definecolor{redstrings}{rgb}{0.9, 0, 0}
\definecolor{graynumbers}{rgb}{0.5, 0.5, 0.5}
\date{\today}
\title{Nefele: Process Orchestration for the Cloud }

\author{
    \IEEEauthorblockN{Mina Sedaghat, Pontus Sköldström, Daniel Turull,
     \\ Vinay Yadhav, Joacim Halén, Madhubala Ganesan, Amardeep Mehta, Wolfgang John} \\
    \IEEEauthorblockA{Cloud Systems and Platforms, Ericsson Research.
    \\\{mina.sedaghat, pontus.skoldstrom\}@ericsson.com}
   
}
\begin{document}


\IEEEtitleabstractindextext{
\begin{abstract}\label{sec:abstract}
Virtualization, either at OS- or hardware level, plays an important role in cloud computing.  It enables easier automation and faster deployment in distributed environments.  While virtualized infrastructures provide a level of management flexibility, they lack practical abstraction of the distributed resources.  A developer in such an environment still needs to deal with all the complications of building a distributed software system.  Different orchestration systems are built to provide that abstraction; however, they do not solve the inherent challenges of distributed systems, such as synchronization issues or resilience to failures. 

\noindent This paper introduces \textit{Nefele}, a decentralized process orchestration system that automatically deploys
and manages individual processes, rather than containers/VMs, within a cluster. Nefele is inspired by the Single System
Image (SSI) vision of mitigating the intricacies of remote execution, yet it maintains the flexibility and performance
of virtualized infrastructures. Nefele offers a set of APIs for building cloud-native applications that lets the
developer easily build, deploy, and scale applications in a cloud environment.  We have implemented and deployed Nefele
on a cluster in our datacenter and evaluated its performance.  Our evaluations show that Nefele can effectively deploy,
scale, and monitor processes across a distributed environment, while it incorporates essential primitives to build a
distributed software~system.
\end{abstract}
}

\maketitle
\IEEEdisplaynontitleabstractindextext

\Pontus{TODO: add Description to all figures to follow ACM rules}
\Mina{What is our final claim? The experiments show that XXX? Easier to build a DS? managed to hide the details well? }
 
\section{Introduction}\label{sec:intro}
Building distributed software systems has always been complicated. In a distributed system, processes are running on
different networked computers, communicating their actions by passing messages over the network, without any notion of a
global clock. In this environment, there are many challenges such as maintaining synchronization, consistency, ensuring
availability, resilience to failures, and traceability. Events occurring in this environment may not appear in the
expected order, partial failures of the system have to be dealt with, and nodes may disagree on the current state of the
system. Many solutions that work well on a single node no longer~apply.

The development of virtualization techniques for the x86 platforms gave rise to cloud computing, where, depending on the
service model, a user rents a set of distributed computing resources for deploying her application. While cloud
platforms provide easy and scalable access to distributed resources, they do not, by themselves, solve the inherent
challenges of distributed software systems mentioned above. To alleviate these issues, cloud providers offer a wide
range of services and products that are engineered to operate in a distributed environment and provide commonly required
functionality such as logging, databases, locking~\cite{burrows2006chubby}, monitoring, and storage. However, the
developer of a cloud application still must carefully craft the application to deal with many of the complications
stemming from the distributed nature of the underlying system.
\begin{figure*}[htbp!]
\centering
\includegraphics[width=.9\linewidth]{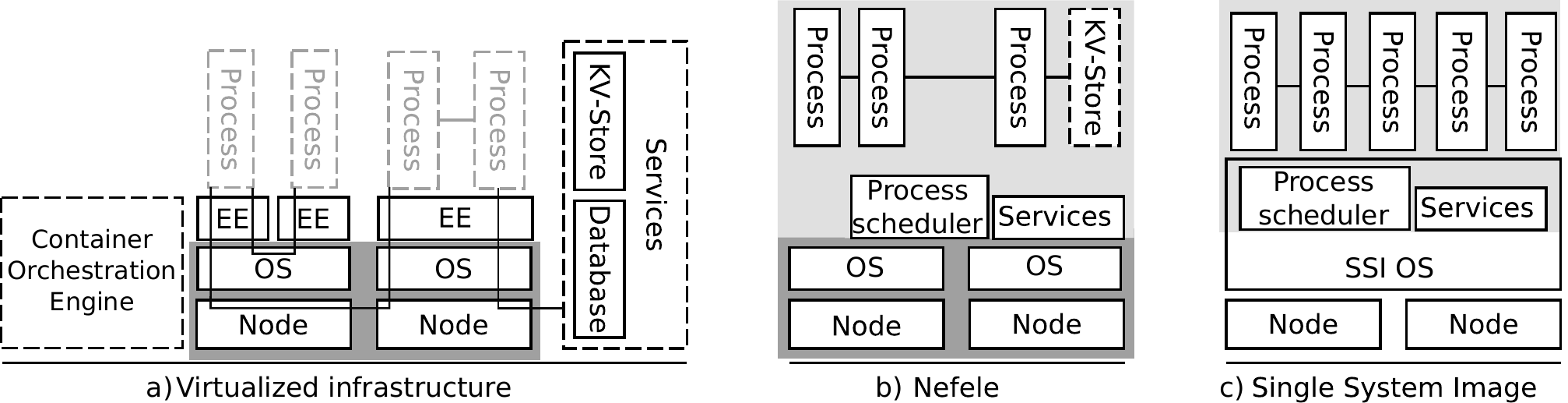}
\caption{Comparison of virtualized infrastructure, Nefele, and SSI.}\label{fig:cloud-nefele-ssi}
\end{figure*}

Multi-core machines share many of the same problems. However, to a large extent, the industry has been able to provide the
illusion of a single core system, to be programmed as a single system. If a similar illusion can be provided for a
multi-node system, the complexities of being distributed could be hidden from the developer and make applications easier
to develop, debug, and operate.

Providing this illusion, or an abstract view of the underlying hardware, is the goal of the Single System Image (SSI)
concept, which has been implemented by many projects since the mid-1980s. However, none of these projects reached large
mainstream adoption, mainly due to low \emph{performance} and lack of \emph{scalability} of the single-node constructs.
Emulation of these constructs, to provide a complete UNIX interface, has fundamental limitations that usually boils down
to the need for synchronization between different nodes. Moreover, many techniques that work well on a single node, do
not work well in a distributed setting. For example, memory sharing between several processes is efficient in a single
node, but not in a distributed environment~\cite{healy2016single}.

Additionally, many of these projects implemented the SSI features on the kernel level, e.g.,\ as a set of patches. This
makes it difficult to keep up with the rapid development of open source kernels.  Finally, to accommodate the
requirements of diverse applications, SSI services typically have implemented the strictest consistency version of,
e.g.,\ a distributed file system, even when most applications actually do not need it. Having to choose the lowest
common denominator to build a generic system can severely impact performance.  In the end, a takeaway from the SSI
efforts is that one size typically does not fit all.

Erlang/OTP~\cite{OTP} (Open Telecom Platform) has shown to be a successful approach to build distributed software
systems. It is a functional language and runtime designed for building distributed, fault-tolerant, and highly available
systems. The core of the Erlang model is to place all computations into strongly isolated processes, sharing no data
between them, and interacting only through asynchronous message passing. OTP extends the basic Erlang language and
runtime with a set of supporting libraries and design principles.

In this paper, we extend our work in~\cite{john2018making}, and we introduce Nefele, a decentralized process orchestration system, that can execute and manage Linux processes on a cluster. Nefele is inspired by the SSI vision of mitigating the intricacies of remote execution and has adopted some of
the Erlang/OTP design principles and mechanisms for dealing with distributed software systems. In Nefele, a developer is
equipped with a set of SSI-like features and programming APIs, in addition to the local OS features in the virtualized
environment. These extra functionalities hide the complexities of process deployment and IPC in distributed
environments, by providing simple programmatic interfaces for processes to deploy, execute, connect, and monitor other
processes.  However, Nefele does not try to hide the fact that the applications are executing in a distributed
environment and therefore does not restrict the developer from using the features that are currently available (or
efficient) on a single node. In Nefele, a \emph{process} is the unit of scheduling and execution, as well as the
endpoint of messaging. A process is a finer unit compared to a container or a VM, providing higher flexibility and
malleability. As in Erlang, dependencies and relationships within an application are defined by the application itself,
rather than using external deployment charts and manifests.

The rest of the paper is organized as follows: Section \ref{sec-2} explains Nefele's design choices and their
rationale. Section \ref{sec-implementation} describes Nefele's architecture and its components, followed by its offered
APIs and interfaces in Section \ref{sec-API}. Section \ref{sec-performanceEval} presents our experimental setup and
discusses the evaluation results. We close the paper with a discussion on related work in Section \ref{sec-relatedwork}
and conclude the paper in Section~\ref{sec-conclusion}.

\section{System design}\label{sec-2} 
In Nefele, we wish to bring the simplicity of the SSI model (from a developer's perspective) into the proven virtualized
infrastructure model, both depicted in Figure~\ref{fig:cloud-nefele-ssi}. In the virtualized infrastructure model
(Figure~\ref{fig:cloud-nefele-ssi}a), the entity responsible for deployment and orchestration, the \textit{Container
  Orchestration Engine}, is often an external system responsible for instantiating and managing execution environments,
e.g.,\ containers or VMs, within which processes run. These processes communicate with each other in a host-to-host
fashion, unless they run in the same execution environment. Processes in an execution environment can use Operating
System (OS) services which are fundamentally not distributed by default. To simplify the task of developing a
distributed software system, services designed to operate in a distributed environment are made available as external
services. Typically, there are multiple such services that provide the same functionality with different
characteristics, allowing the developer to chose the most suitable for a task.

In Figure~\ref{fig:cloud-nefele-ssi}c the SSI model is depicted. In the SSI model, the \textit{Process Scheduler}
performs a similar role to the container orchestration engine, however, it directly deploys and manages processes rather
than the execution environment.\@ These processes communicate with each other directly without any notion of the
underlying hosts. The SSI OS inherently provides functionalities that are needed for developing distributed software
systems, e.g.,\ an Inter Process Communication (IPC) system or a consistent distributed file system. Typically, the user
cannot make any choice regarding these functionalities, and she is bound to what the system provides. The system, in
turn, must implement the most strict form of these functions in order to support the most strict application
requirements, for example by using a strongly consistent distributed file system rather than one that is eventually
consistent.

Nefele aims for a hybrid model, inheriting useful aspects from each. Figure~\ref{fig:cloud-nefele-ssi}b shows the design
of Nefele, with the shaded backgrounds highlighting the overlaps with the other models. Nefele adopts the ideas around
shared process space, IPC, process placement, and a restricted set of OS services from SSI, and enhances it with the
ability to see and interact with the underlying single-node OS, and utilize the currently available tools from the
virtualized infrastructure. Thus, processes managed by Nefele can benefit from both node's local functionality, as well
as distributed SSI services.

\subsection{Design choices}\label{sec-2-1}
There are several design choices to make when combining two approaches of the virtualized infrastructure- and SSI model.
The level of \textit{transparency} of an SSI system is often defined in terms of what aspects of the full system are
aggregated into a single view~\cite{healy2016single}.  The following is the list of design choices made in Nefele,
combining aspects of the SSI, virtualized infrastructure, and Erlang/OTP models:

\subsection{Single system image}
\noindent \textbf{Single process space} (abstract view of a global process table): This is at the core of both the
Erlang and SSI models and can simplify programming, as the developer no longer needs to bother about the node running
the process. For this reason, Nefele assigns a cluster-wide PID (the NPID) to each user process, which is used for
interacting with other Nefele processes in the system through the Nefele interface. Each process also retains its normal
POSIX process ID which it can use with, e.g.,\ existing Linux system calls. This lets the developer to take advantage of
both the SSI- and virtualized infrastructure view without enforcing either.

\noindent \textbf{Single IPC space} (global inter-process communication): The benefit of a single process space
becomes more evident when processes can send messages to each other using the PID, regardless of the location.
Nefele provides a process-to-process messaging system, where messages are sent to a process using its NPID. Other
identities such as registered names may also be used as communication endpoints. The process-to-process approach is
usually easier to use than the host-to-host communication typically offered in the virtualized infrastructure model,
where there are a plethora of different solutions for solving various issues~\cite{newstack}.

\noindent \textbf{Single root} (globally shared filesystem): Having a fast, local, filesystem as in the virtualized
infrastructure is useful, e.g.,\ to store intermediate results, logs, state, etc. On the other hand, a
distributed filesystem as in the SSI model is a simple mechanism for sharing large amounts of data between
processes. To achieve this, Nefele provides both a local and a distributed file system. This gives the developer the
freedom to choose which one is appropriate for each task, by using different filesystem directories,
thus avoiding unnecessary synchronization.

\noindent \textbf{Single I/O space} (global access to locally connected devices): In the SSI model peripheral devices
such as printers, block devices, and GPUs that are connected to a node should be accessible from any other node.  In the
virtualized infrastructure model such devices are typically only accessed locally, through a virtualization
layer. Access to such devices is usually managed by the orchestration engine, where the user explicitly requests a
device, and the job is placed on a node where such a device is available.  Since providing transparent remote access to
a peripheral device, such as an hardware accelerators like GPUs or FPGAs, would have negative performance implications,
we only allow access to nodes' local devices, as done in virtualized infrastructures.  In Nefele the need for an
accelerator would be expressed as a resource requirement and the process will be placed on a node where such a device is
locally available.

\noindent \textbf{Distributed shared memory (DSM)} (one global, shared, memory space): The performance of DSM is still
far from the performance of local memory, e.g.,\ a local page fault takes in the order of 0.1 µsec to resolve whereas a
remote memory copy using a network protocol like~RDMA takes on the order of 10 µsec. Therefore, Nefele
only offers local memory to its processes. Given the amount of RAM memory available in modern servers, we
believe that this restriction will not significantly impact most applications. 

\noindent \textbf{Process checkpointing and migration} (Pausing and migrating processes between nodes): This is a useful
feature implemented both in SSI systems as well as in virtualized infrastructure environments \citep{criu}. However, in
virtualized infrastructure environments it is typically provided on the level of the execution environment. Due to its
usefulness in balancing loads, adapting to usage patterns, etc., we plan to provide such features in Nefele as well,
although they are not yet implemented.

\subsection{Virtualized infrastructure}
The virtualized infrastructure model offers several useful features, that we aim to maintain and benefit from:

\noindent \textbf{Resource control and isolation} is essential to support multi-tenancy where different user processes
must be well isolated to reduce the risk of user processes interfering with each other. The extra isolation is primarily
required to prevent one process from monopolizing a certain resource and, therefore, causes others to starve or
crash. This is the reason Nefele not only supports resource control between tenants but also between their own
processes.

\noindent \textbf{The notion of discrete physical machines:} Allowing a process to see which node it is running on makes
it possible to take advantage of efficient local OS features such as shared memory, storage, and low-latency
communication within that node. Being able to distinguish remote nodes and their location is also necessary for
controlling availability and redundancy, e.g.,\ by replicating processes on different nodes in other racks or
clusters. Therefore, Nefele does not try to hide this fact and supports both a cluster-wide view as well as a local
view.

\noindent \textbf{Resource bundling} in the form of VM or container images allows the user to package all the
dependencies of an application (executables, libraries, configuration files, etc.) into a consistent environment. These
resource bundles can be transferred and deployed on different, heterogeneous nodes, and ensure that applications can run
correctly regardless of differences of nodes environments. This concept has proven extremely valuable and therefore is
used in Nefele as well.

\noindent \textbf{External services} that are commonly used, such as distributed logging, are either provided as
built-in Nefele features or made available as services. The challenge here is to determine which services should be
built-in and which ones made available as services. Once a service is built-in, it must be available everywhere and be
performant enough for most user applications.

\subsection{Erlang/OTP}

Finally, we have adopted a set of functionalities and design principles from Erlang/OTP:

\noindent \textbf{Processes as actors:} The \textit{Actor model}~\cite{actormodel} is a design pattern that has proven
successful for implementing concurrent systems. In this model, an actor is a computational entity, e.g.,\ a process,
executing concurrently with other actors. Each actor has a local state and an address, which it uses to interact with
other actors through message passing. Message passing prevents a multitude of complex issues related to sharing
state. An actor may also spawn additional actors.

This model is supported by Nefele where each process (implemented as a Linux process) has an address (the NPID) and a
mailbox for receiving and sending messages, and the ability to spawn additional processes. While we advocate following
the actor model there are cases where breaking the model makes sense. For this reason, we do not enforce the model but
allow the developer to choose.
  
\noindent \textbf{Fault tolerance:} Nefele adopts one of the key fault-tolerance mechanisms from Erlang/OTP, i.e.,
supervision trees~\cite{armstrong03}. Supervision trees consist of processes responsible for monitoring worker processes
and restarting one or more of them should one crash. This concept moves the responsibility of handling errors from the
worker process to an external entity and provides a clean separation of fault handling from the process' functionality.
The supervisor can run on a different node, allowing the system to tolerate hardware errors when a process fails. In
addition, the developer no longer need to trust the process to heal itself. It reduces code complexity, as custom logic
for failure recovery is no longer needed.

\section{System implementation} \label{sec-implementation} 
\begin{figure}
\centering
\includegraphics[width=.9\linewidth]{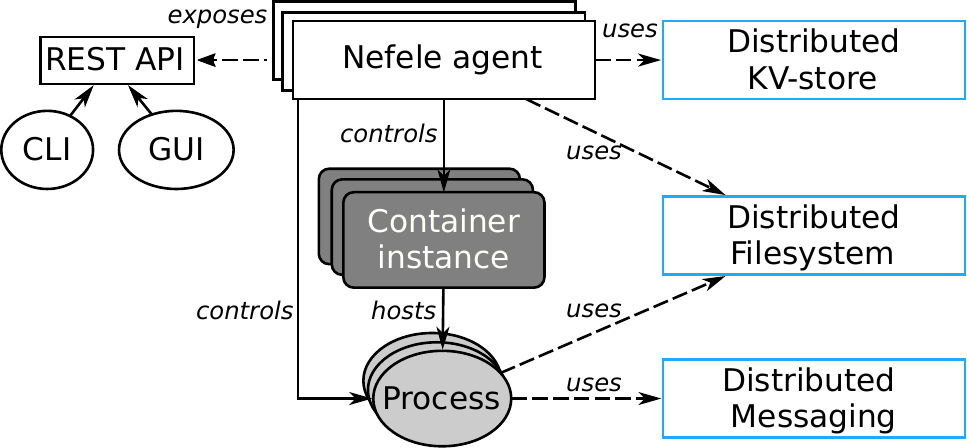}
\caption{Nefele architecture, where each node runs a Nefele agent which in turn manages multiple container instances, each hosting multiple processes.}\label{fig:nefarch}
\end{figure}
We have implemented Nefele to simplify building cloud native (distributed) applications, by realizing the design choices
laid out in the previous section. As such, we must manage and control four different layers: 1) the cluster, i.e., a
collection of several nodes, 2) the compute nodes, 3) the tenant, where multiple-tenants share a single node, and
finally 4) the processes.  The implementation consists of a distributed \textit{control-plane} and a distributed
\textit{data-plane}.  The control plane is responsible for managing different aspects at each of the four different
layers, whereas the data-plane is responsible for creating the execution environment for user processes and executing
them. So far, we have focused on a cluster of bare-metal nodes and a container-based execution environment, however,
Nefele can in principle run on a cluster of virtual machines or use virtual machines as the execution environment
instead of containers.

\begin{figure*}[h]
\centering
\includegraphics[width=.9\linewidth]{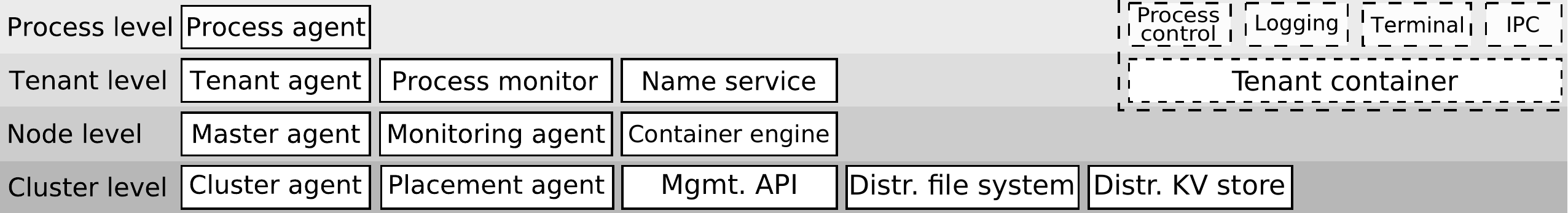}
\caption{The Nefele control-plane modules acting at different system levels. Dotted boxes represent data-plane components at the respective levels.}\label{fig:erlang-runtime}
\end{figure*}

Nefele architecture, its high-level components, and their interactions are depicted in Figure~\ref{fig:nefarch}.  Each
node runs an instance of the Nefele control-plane agent, and a group of Nefele agents forms a distributed control-plane.
The Nefele agent on each node is responsible for creating and controlling containers belonging to different tenants, to
spawn and control processes spawned within those containers, and more importantly to coordinate and interact with other
Nefele agents to provide a common process and IPC space. The Nefele agent is mostly implemented as a distributed Erlang
application, with some additional helper components running outside of Erlang. Nefele agents use a distributed KV-store
and a distributed filesystem used to store, e.g.,\ process state and container images respectively.  The distributed
file system can also be used by \textit{user processes} to, e.g.,\ persist state or share data. The user processes have
access to a distributed messaging system for communicating with each other. Finally, each Nefele agent exposes a
REST-based management interface for users to manage their processes, images, etc.

Figure~\ref{fig:erlang-runtime} shows the control-plane and data-plane components, acting on different layers.  To
better understand the Nefele implementation, we describe the required functionalities of each layer, followed by the
responsible component~specifications.

\subsection{Cluster-level management}
\label{sec-4-1}
A cluster consists of a collection of nodes whose resources (CPU, memory, devices, etc.) are to be aggregated and
controlled. Often, adding resources to a cluster requires adding more management capacity. Existing cluster managers
have upper limits on the number of nodes and applications they can support (e.g., Kubernetes can support up to 5k nodes
and 300k containers per cluster manager). After reaching the limit, adding more nodes requires setting up a new cluster
manager which is often complex and time-consuming. In Nefele, each node is part of the distributed control-plane and
contributes to the management capacity, as more nodes join.

At the cluster-level, we have five different control-plane components, each running in all the nodes.

\noindent \textbf{Cluster agent}\hspace{3pt}
The cluster agent is a control-plane component responsible for discovering and structuring nodes to form resource groups
in the cluster. It is also responsible for performing health checks and detecting node failures.  The cluster agent
interacts with other cluster agents to maintain a list of neighboring nodes.  Groups of cluster agents automatically
structure themselves into hierarchical resource groups.  The groups can dynamically adapt to the changes in the
environment, e.g.,\ if a node fails or is removed for maintenance a group may shrink or get merged with another
group. The cluster agent is implemented as a self-organizing membership management system, based on
SWIM~\cite{das2002swim, du2018self}.

\noindent \textbf{Placement agent}\hspace{3pt}\label{sec-6-1-4}
The placement agent is a control-plane component, implementing a fully decentralized scheduler responsible for placement
and scheduling of processes within Nefele. It receives requests to spawn processes and it then processes them
asynchronously. Each request is a collection of identical tasks which each demands a certain amount of resources, and if
deployed starts a process.  The placement agent follows the \textit{feasibility and ranking} principle, it first
performs a feasibility check to identify the suitable nodes, and then scores them according to a preference order. As
part of the feasibility check, the placement agent queries other nodes in the cluster for resource availability and
checks whether they can deploy the whole request, i.e., all tasks, or a fraction of it, i.e., a subset of tasks. The
process of scoring and ranking the potential locations to place a task can be computationally expensive. This scoring
process is parallelized by having each node calculate its own score based on its available resources, the risk of
resource stranding if the request is accepted, and the risk of over-subscription given the current load
fluctuations~\cite{sedaghat2016decentralized}.  This distributed approach reduces the probability of requests queuing up
and improves scheduling time and throughput.

\noindent \textbf{Management API and application images}\label{sec-6-2-1}\hspace{3pt}
The management API is provided by another control-plane component. It exposes a REST interface that lets users control
different aspects of Nefele, such as starting and stopping applications, connect to running applications, upload
and download files to and from applications, check the status of running applications, register and authenticate users,
and to upload application images. Applications in Nefele are packed as system images, just like OCI images~\cite{oci-image-format}. 
Once uploaded, the system images are placed on a distributed file system mounted on each node. These
images are then mounted as the root filesystem on-demand, as containers for hosting user processes. The user can
choose to run in either development or production mode. In the development mode, the image is mounted with read/write
permissions and any modifications are stored using an overlay file system. This is to simplify the development process,
making it easy to modify the system while testing.  In production mode images are mounted read-only, leaving only the
shared file system and certain file system paths writable.

\noindent \textbf{Distributed file system}\hspace{3pt}
The distributed file system is another part of the control-plane, used to distribute data among the nodes in the
cluster. Its primary use is to distribute the user's application images. Currently, we use Gluster~\cite{gluster} as
the distributed file system for both control-plane and offered to a user applications, so that processes may persist
data and/or share data with each other. We aim to offer support for other file system types, so that users can choose
the one appropriate for their application.

\noindent \textbf{Distributed KV-store}\hspace{3pt}
The distributed KV-store is a control-plane component used to distribute and persist control-plane meta-data among the
different control-plane nodes. This includes data about running processes, registered users, monitoring data, etc. We
currently use Redis~\cite{Redis} in this role, however, there are many other feasible implementations.

\subsection{Node-level management}
Individual nodes in the cluster, whether physical nodes or virtual machines, run three different control-plane
components; the master agent, the monitoring agent, and the container engine.

\noindent \textbf{Master agent} \hspace{3pt}
The master agent coordinates the local and remote components. It relays requests between local modules, e.g.,\ passing
process spawn requests to the placement agent to be allocated.\@ It  is also responsible for passing requests to
components in remote nodes via the remote master agent. It also interacts with external, non-Erlang, components, e.g.,\
by forwarding requests to create a container to the external execution environment agent.

\noindent \textbf{Monitoring agent}\hspace{3pt}
The monitoring agent gathers resource consumption data from the local node and containers. This data is primarily used
by the placement agent to assess whether or not it has enough resources to accept a task.

\noindent \textbf{Container engine}\hspace{3pt}
The container engine is responsible for creating, configuring, and controlling tenant containers for user processes to
execute in. Rather than implementing our own container engine, we choose one of the existing open source
implementations. There are many of them to choose from, e.g., \texttt{containerd}\footnote{https://containerd.io} or
\texttt{CoreOS rkt}\footnote{https://coreos.com/rkt}. For Nefele, we decided to use the built-in container functionality
of \texttt{systemd}, called \texttt{systemd-nspawn}. We chose \texttt{systemd-nspawn} for its convenient D-Bus interface
which allowed us to implement the interactions like asynchronous D-Bus calls, which fits well into the Erlang message
passing design.

\subsection{Tenant-level management and execution}
\label{sec-4-2}
In Nefele, a tenant consists of one or more users that can share the allocated resources. Resources and applications
belonging to different tenants should be isolated from each other to prevent information leakage and performance
interference between tenants. On the compute side, containers are the means to provide that isolation, where different
resource views are implemented through various Linux namespace functions and resource usage control through Linux
control groups (cgroups). To prevent messages from leaking between tenants, the tenants are deployed in different
network namespaces and the underlying messaging system separate tenants traffic into different virtual networks.

To provide the single process space from the SSI model, a tenant application is not modeled as a collection of
independent containers, but as a single application image whose executables can be transparently deployed over multiple
nodes. In each node where one or more executables should be spawned, a single container and the associated networking
for that tenant is created. Coordinating this over multiple nodes is the job of the tenant-level parts of the
control-plane.  The tenant-level control-plane modules also provide the services needed for each tenant on a node.

At the tenant-level, there are three control-plane components; the tenant agent, the process monitor, and the name
service agent.  The tenant has also a data-plane component, the execution environment itself.

\noindent \textbf{Tenant agent} \hspace{3pt}
A tenant agent is started by the master agent when the first process belonging to the tenant is placed on the node.  The
tenant agent triggers the creation of a tenant container instance and is responsible for managing that instance. As the
tenant's application spreads over multiple nodes, multiple instances of the container are created, each on a different
node, managed by a different tenant agent. Different tenant agents are then communicating with each other to create the
view of a single process space over a cluster.

The tenant agent receives requests from processes running inside the tenant container through a UNIX domain socket
shared with the processes inside the container. These requests may, for example, be requests to spawn additional
processes, send signals to existing ones, or list all running tenant processes. The tenant agent replies to those
requests by interacting with other local or remote modules. This may involve, e.g.,\ spawning a process in the local
tenant container at the decision of the placement agent, request a tenant agent on another node to emit a SIGKILL signal
to a tenant process running there, or obtaining the list of running tenant processes from the distributed KV-store.

\begin{figure}[tb]
\centering
\includegraphics[width=.9\linewidth]{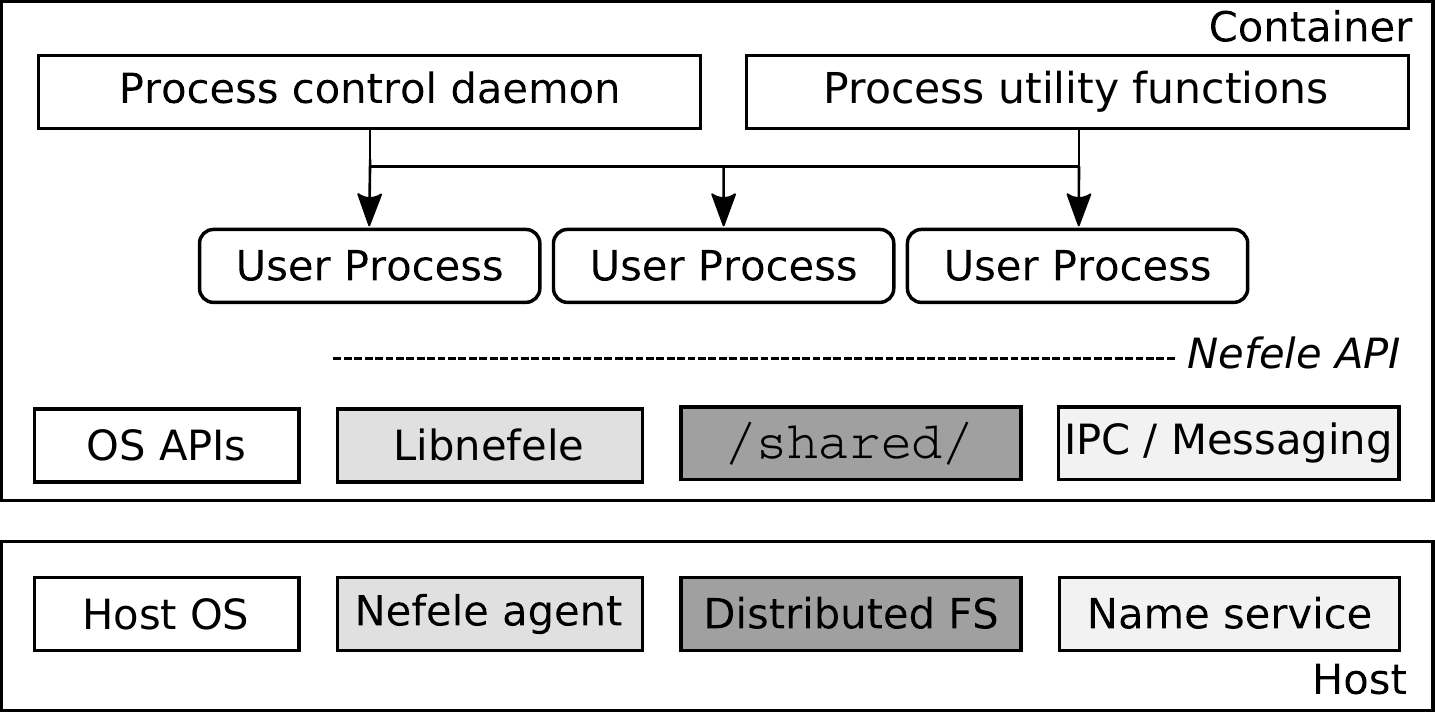}
\caption{User and system processes in a tenant container, and their relationship to control-plane functions.}
\label{fig:host-cont}
\end{figure}

When the tenant agent operates in the local tenant container, e.g.,\ on its own local processes running inside the tenant
container, it interacts with the \textit{process control daemon} inside the container (see Figure~\ref{fig:host-cont})
which executes the actual commands. These interactions are done through the Process monitor.

\noindent \textbf{Process monitor} \hspace{3pt}
The process monitor connects to the system D-Bus instance running \textit{inside} the tenant container and interacts
with various data-plane modules there. Its primary function is to monitor the creation and termination of user processes
by subscribing to event notifications. It is also used to call RPC functions in the other modules, e.g.,\ to spawn a
process, allocate pseudo-terminals, and emit signals.

\noindent \textbf{Name service} \hspace{3pt}
The name service is the control-plane module of the distributed messaging system and provides a mapping between user
process addresses and the names they have registered. It also provides mechanisms for creating and addressing groups of
processes, e.g.,\ placing different ``webserver'' processes into a single group. A name service instance runs for each
tenant and distributes the mappings among name service instances on each of the nodes where the tenant is present. The
tables with the name-to-address mappings are exposed inside the tenant container as a read-only database.

\noindent \textbf{Tenant container}
The tenant container instance is our first data-plane module, shown in Figure ~\ref{fig:host-cont}. It is created with
certain resource control settings, its own network namespace, and a set of paths mounted inside it. These paths include
the distributed file system (in \texttt{/shared}), the name service database, and the UNIX domain socket used to
communicate with the rest of the control-plane.

\begin{table}[!t]
  \renewcommand{\arraystretch}{1.3}
  \caption{Process utility functions.}\label{tab:procmgmt}
  \centering
  \begin{tabular}{lp{6cm}}
    \toprule
    Monitoring   & \multicolumn{1}{p{6cm}}{\raggedright A process can register to get notified when another process terminates.} \\
    Listing      & \multicolumn{1}{p{6cm}}{\raggedright A process can obtain a list of running  processes  (similar to \emph{ps}).} \\
    Signaling    & \multicolumn{1}{p{6cm}}{\raggedright A process can send POSIX signals (e.g., SIGINT) to another process.}\\
    Standard I/O & \multicolumn{1}{p{6cm}}{\raggedright A process can obtain the output  (\emph{stdout}, \emph{stderr}) or connect 
                   to a per-process pseudo-terminal for I/O.}\\
    Logging      & \multicolumn{1}{p{6cm}}{\raggedright A process can obtain the logs of one or more processes.}\\
    IPC          & \multicolumn{1}{p{6cm}}{\raggedright The control-plane also manages name mappings for 
                   the data-plane IPC system, discussed in Section~\ref{sec-4-3-2}.}\\
    \bottomrule
  \end{tabular}
\end{table}

\subsection{Process-level management}\label{sec-4-3}
The Nefele control-plane is responsible for managing all tenants' processes. It provides functionality and APIs for
process deployment, scaling, monitoring, and signaling. Through the API, the developer can dynamically spawn one or more
processes in a distributed fashion and obtain a process handle (Nefele PID or \emph{NPID}). In the function call, the
developer can specify the process resource and location requirements such as affinity/anti-affinity, CPU/memory
requirements, and configure the environmental variables. Nefele allows spawning a single process (\texttt{spawn}), a set
of N identical processes (\texttt{nspawn}), and a set of different processes with different requirements
(\texttt{cspawn}).  Spawning processes in groups rather than one by one allow us to better optimize the placement, it
also reflect common request patterns in cloud applications, e.g., map-reduce.

The control-plane is also intended to provide additional functionality related to process checkpointing and restoration
as well as migration, these are however not implemented as of writing.  Other process management functions are
summarized in Table~\ref{tab:procmgmt}.

\noindent \textbf{Process agent}\hspace{3pt}
The only control-plane component at this level is the process agent, an Erlang process that \emph{shadows} a user
process. It is responsible for the initial synchronization with the started process, which is done through a shim
library, preloaded before the process starts. After initialization, it monitors the process and notifies the Tenant
agent when the process is terminated.

\noindent \textbf{Process control daemon}\hspace{3pt}
In the data-plane, each container runs a process control daemon which spawns user processes, configures the resource
control for the process, sets its environment variables, monitors its liveness, and records the exit status of the
process. Each user process must synchronize with its corresponding process agent in the control-plane, to exchange
identifiers and initialize the messaging system. To enforce that each process does this, we use \texttt{LD\_PRELOAD}
~\cite{kerrisk2010linux} to load a shim library that is executed in the process before continuing to the process entry
point, e.g.,\ the \texttt{main()} function.  The control-plane synchronization can be done relatively quickly, and
processes, in general, can start in a matter of milliseconds (see measurements in Section~\ref{sec:performance}).

Rather than implementing our own process control daemon, we looked at different open source implementations and
currently, a \texttt{systemd} instance running as PID 1 in the container is our process control daemon. This currently
restricts Nefele to only use systemd-based application images, however, implementing a generic daemon providing the same
functionality would solve this issue. We also make use of the \texttt{systemd} functionality for redirecting the
standard I/O file descriptors of a process upon startup. Typically, they are redirected to write to the logging system
(currently \texttt{journald}~\cite{journald}), but they can also be redirected to the Terminal service if needed.

\noindent \textbf{Logging service}\hspace{3pt}
The logging service implements a realtime streaming log functionality, where a process can request to receive realtime
logs from processes running in the system, on any node, with certain filters if requested (e.g., filtered by NPID). The
logging service can also export logs to an external logging database for later viewing or processing.

\noindent \textbf{Terminal service}\hspace{3pt}
The Terminal service also runs in the data-plane and lets other processes either retrieve the standard output and error
streams of a process, or to interact with it through a pseudo-TTY, making the terminal service similar to an SSH server
combined with \texttt{screen\footnote{https://www.gnu.org/software/screen/}}.

\noindent \textbf{Messaging system}\label{sec-4-3-2}
The messaging system runs in the data-plane to implement a single IPC space. It is implemented as a distributed,
broker-less, messaging system that supports several different communication patterns and identities. Before a user
process is executed, the messaging system creates a mailbox for it, this mailbox starts collecting messages received
from other processes and notifications from the control-plane.  To send messages to different processes there are
several different addressing options. The first option is to use the process handle, (i.e.,\ the NPID), as the
destination of a message. This is particularly useful for related processes, e.g.,\ from parent to its child.

However, to send a message to an unrelated process (i.e.,\ not a parent or child) one must then first find out its
NPID.\@ To simplify this Nefele allows processes to register as \textit{services}, using a location independent
numerical service identifier or a name. The sending process can now simply address the message using, e.g.,\ the name
``webserver'' or the identifier ``80'' if the receiving process has registered for this name or ID.\@ These DNS-like
process names and identifiers are automatically distributed to tenants in different nodes and mapped into the memory
space of the Nefele processes. This provides fast service lookups and a mechanism for processes to list available
services.

In addition to simple point-to-point messaging, Nefele offers several other messaging mechanisms, for example, automatic
failover between services of the same type (e.g.,\ if there are two ``webserver'' processes). Location independent
names/identifiers combined with automatic failover makes it easier to handle typical failover or migration scenarios, at
least the networking aspects.  Other messaging mechanisms that can be used are different types of publish/subscribe
trees and multi-casting. \Pontus{Add something here about the external proxy functionality!}

\section{APIs and user interfaces}\label{sec-API}
Nefele provides APIs for different languages (C, Go, and Python), however, the developer is free to implement our
protocol (based on Protocol Buffer serialization~\cite{protobuf}) and directly communicate with the local Nefele agent.
Using these interfaces, several useful mechanisms can be implemented, with the typical application consisting of
processes passing tasks around as messages and spawning new worker processes to handle jobs in parallel.

Supervisors, a fault tolerance mechanism, can be implemented using primitives for process \textit{\textsf{spawning}},
\textit{\textsf{monitoring}}, \textit{\textsf{signaling}}, and \textit{\textsf{messaging}} (using functions shown in
Listing~\ref{functions}).  In this case, one spawns a process that acts as a supervisor, which in turn spawns and
monitors other processes. These processes may be workers or more supervisors. If a supervisor is notified that one of
its children has crashed, it can restart all its children or just the crashed one, depending on the strategy and
dependencies among the children. This is a mechanism successfully used for constructing fault-tolerance in Erlang
applications. The supervisor approach can be further extended to support scaling of its associated processes, e.g.,\ the
supervisor collects the status of its processes and dynamically adjust their load by spawning new children or killing
existing ones. In this case, the automatic failover and load-balancing features in the messaging system simplify the
issue of directing traffic to / away from new/old processes. The communication between the processes remains intact,
without the need to manually re-establish new connections, when the processes are restarted or redeployed.

Synchronizing processes and their startup is easy using the messaging system. Processes can \textit{\textsf{wait}} for
incoming messages before taking a particular action or wait for other processes to be started before continuing with
their own initialization. By \textit{\textsf{registering}} service names ordering the startup of dependent processes is
even easier. Service names combined with supervisors allows one to design dependency trees with ordered startups,
somewhat similar to how services are organized in the systemd service manager.

\begin{lstlisting}[language=c, stringstyle=\ttfamily, label=functions, caption=Select functions from the Nefele API.]
// spawn a process
npid_t* nefele_spawn(char* path, char** env, ...);
// monitor a process
int nefele_monitor(npid_t* npid);
// signal a process
int nefele_kill(int signal, npid_t* npid);
// message a process, variable destination type
int nefele_message(void* dest, uint8_t* buf, size_t len);
// wait for a process to be available
int nefele_wait(void* ident);
// register a service name
int nefele_register(char*  name);
\end{lstlisting}

To allocate shared resources for a set of processes, one can spawn a Nefele process that spawns local processes using
the traditional POSIX \texttt{fork()} or \texttt{exec()} calls. In this case, resources reserved by the Nefele process
will be shared among its children, which are guaranteed to run on the same node. This can also be a useful mechanism for
spawning and controlling applications that one does not wish to modify to support the Nefele APIs, the parent Nefele
process can in these cases act both as a proxy and a supervisor for the legacy processes that it manages.

In addition to the protocol, libraries, and management APIs we provide different CLI tools for managing the applications
and manually controlling processes, called \texttt{nefele} and \texttt{nef} respectively.  In the case of application
management, we mimic the \texttt{git} and the \texttt{docker CLIs}, and provide commands to, e.g.,\ save changes to an
image, to push these changes, and to start an application. For process control in a cluster, we mimic traditional UNIX
tools and provide commands to, e.g.,\ spawn processes, monitor them, list running processes, and send UNIX signals (see
Listing~\ref{testy} for examples).\\

\begin{lstlisting}[language=bash, stringstyle=\ttfamily, label=testy, caption=Nefele CLI for image and process management.]
home:~$ nefele commit           # commit changes
home:~$ nefele push             # push changes
home:~$ nefele start            # start application
home:~$ nefele connect app      # SSH to application
app:~$ A=$(nef spawn /bin/prog) # spawn and store NPID
app:~$ nef monitor $A           # monitor process
app:~$ nef ps                   # list running process          
app:~$ nef killall -9 prog      # kill matching 'prog*'
\end{lstlisting}

\section{Performance evaluation}\label{sec-performanceEval}
In our previous work \cite{nefeleicacdemo}, we have demonstrated how the Nefele APIs can
be used to simplify and construct a fault-tolerant, elastic, distributed application for an IoT use-case. 
In this section, we want to focus on evaluating Nefele's performance using a cluster of 15 nodes, each with 16 HT-enabled Intel Xeon processors and 125 Gb of RAM, interconnected by a 10 Gbit/s Ethernet network.  We synthetically generate different workloads (each a series of
requests to deploy jobs) to stress the system and to monitor system behavior in different situations such as high
request arrival rates, different request sizes, and different cluster loads. Each request is a collection of identical
tasks, each demanding a certain amount of resources and starting a
\texttt{stress-ng}\footnote{\url{https://kernel.ubuntu.com/git/cking/stress-ng.git/}} process if
accepted. \texttt{Stress-ng} generates a configured load of CPU and memory usage, and runs for a certain amount of time
(the request's execution time). These values are generated from different normal distributions.

In our experiments, requests arrive following a Poisson distribution, where the interarrival rate is calculated given
the arrival of requests over the simulation time. To impose different types of load on both the control- and data-plane,
we change the distribution parameters for the normal- and Poisson distributions. Each Nefele node can act as an
admission node for resource requests. An admission node receives a request, initiates a placement process, and finally
deploys it over one or several nodes, if accepted. In our experiments, we control which admission node(s) receive
requests in order to evaluate how the distribution of scheduling and management decisions affect the overall throughput
and the time to place them.

In the following experiments, we evaluate different aspects of Nefele; 1) scheduling performance under CPU load, 2)
scheduling performance depending on request size, 3) scheduling throughput of a single admission node, 4) scheduling
throughput of multiple admission nodes, and finally, 5) efficiency of the process monitoring system with regard to
different process spawning mechanisms.

\begin{figure*}[htbp!]
  \begin{subfigure}{8cm}
    \includegraphics[width=0.9\textwidth]{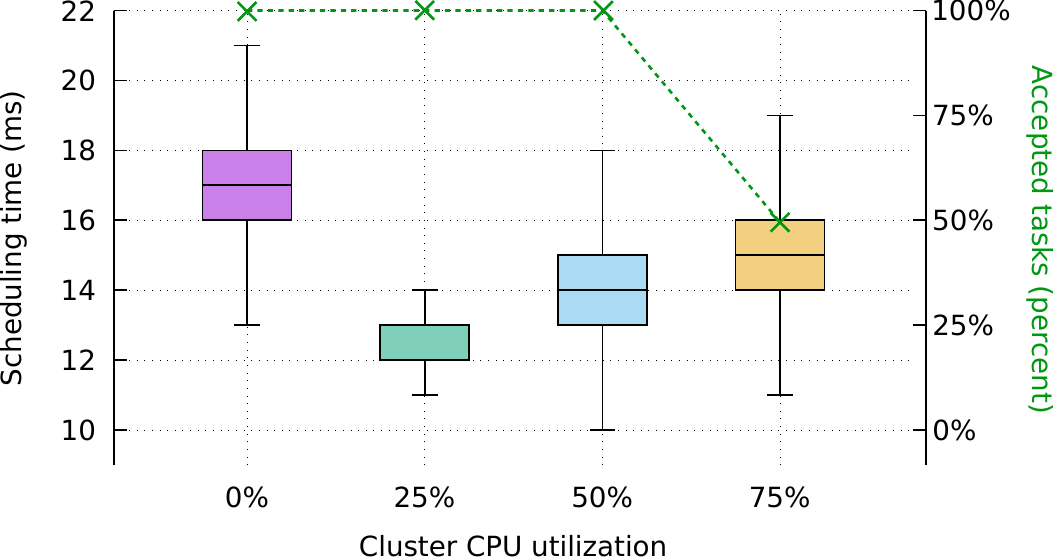} 
    \caption{Single admission node, varying total cluster CPU load.}\label{fig:ClusterLoad}
  \end{subfigure}
  \begin{subfigure}{8cm}
        \includegraphics[width=0.9\textwidth]{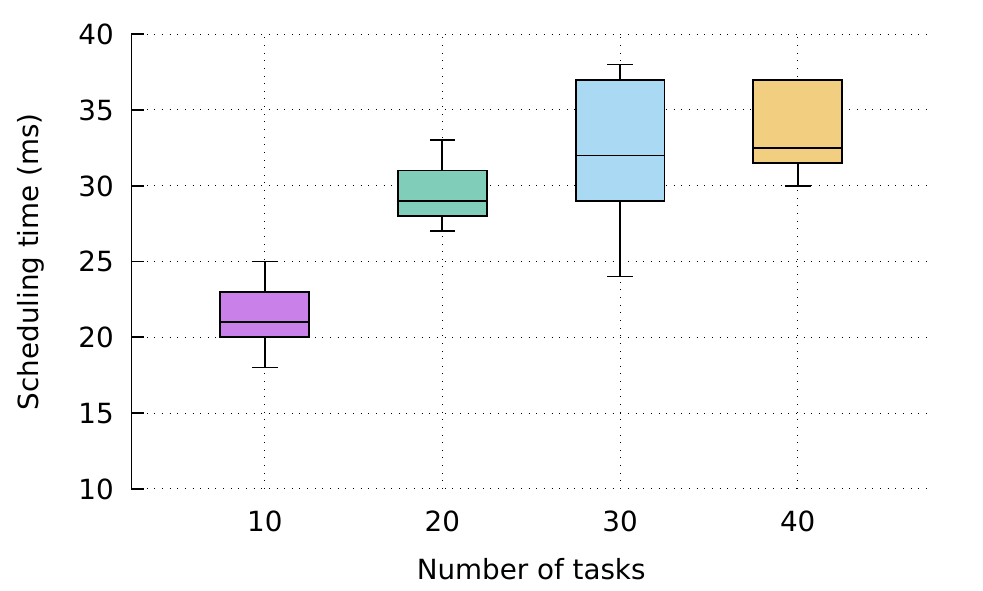} 
        \caption{Single admission node, varying number of tasks.}\label{fig:Tasks_n1}
      \end{subfigure}
      \begin{subfigure}{8cm}
        \includegraphics[width=0.9\textwidth]{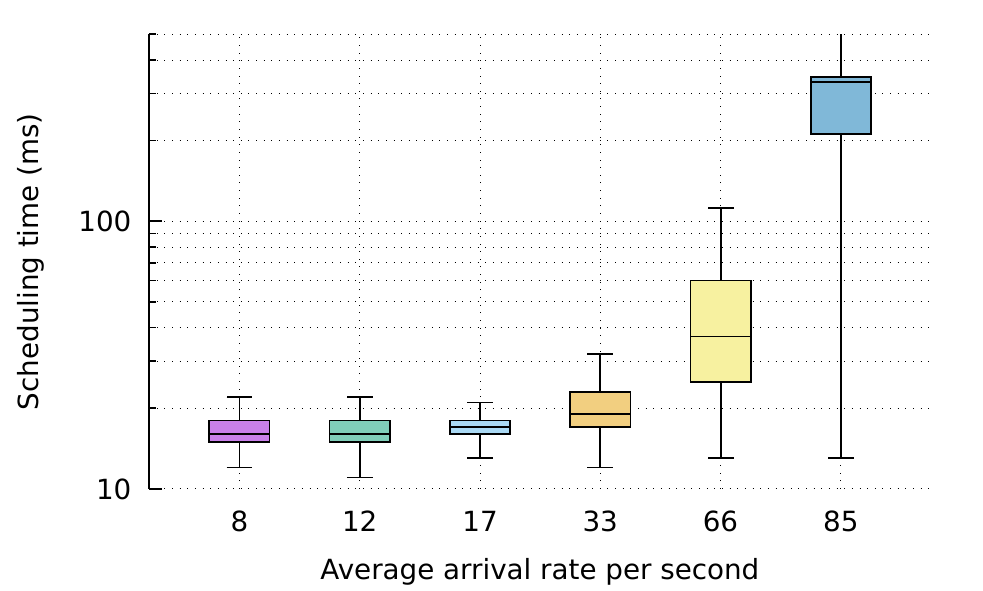} 
        \caption{Single admission node, varying arrival rate.}\label{fig:n30_highThroughput}
      \end{subfigure}
      \begin{subfigure}{8cm}
        \includegraphics[width=0.9\textwidth]{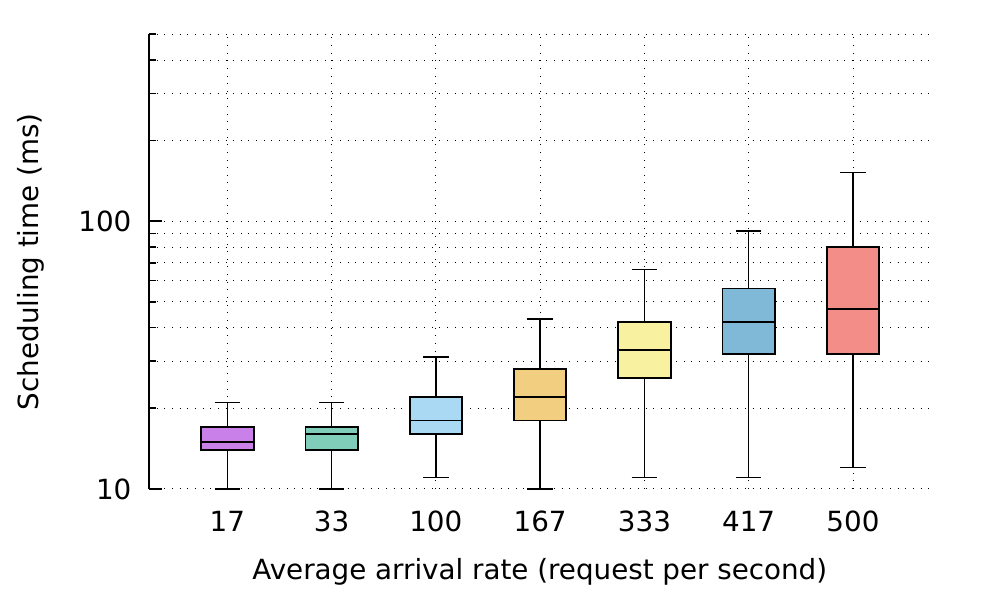} 
        \caption{Multiple admission nodes, varying arrival rate.}\label{fig:n60_highThroughput}
      \end{subfigure}
      \caption{Scheduling time in different experiments.}
\end{figure*}

\subsection{Placing processes in the cluster}\label{sec-5-2}
First, we evaluate Nefele’s distributed process scheduler, and in particular the \textit{scheduling time}.  The aim is to
understand how our design decisions such as distributing scheduling requests and sharing container runtime affect the
scheduling time. We define the \textup{scheduling time} as the time from when a request is submitted to Nefele until the
process is deployed and ready to execute. This time is the sum of the time that the request is processed by the Nefele
scheduler and the deployment time for the tasks (spawning the processes).

We have decided not to include the \texttt{get task} time (as defined in~\cite{sparrow}), the time to transfer the
application to the node, in our scheduling time definition. The \texttt{get task} time depends on the image size and it
is unavoidable as the scheduler needs to ship the task to the worker node. As this time is independent of the scheduling
algorithm, we have decided to not include it. In addition, Nefele shares a container between multiple tasks of the same
tenant, therefore reducing the number of image transfers.

In our first experiment, we evaluate the scheduling time depending on the background cluster CPU load. The intuition is
that it takes longer to find and allocate resources when the cluster CPUs are highly loaded compared to lightly loaded
clusters. We generate a background load by placing \texttt{stress-ng} processes on each node, each running in a tenant
container, and then start issuing requests.

The results, shown in Figure~\ref{fig:ClusterLoad}, demonstrate that for the requests of the same size, Nefele can
maintain an acceptable response time between $\approx$ 14 ms to 20 ms, regardless of the background cluster CPU load. As
shown in the figure in green line, at high cluster CPU loads, Nefele starts rejecting the requests, without
significantly impacting the scheduling time. This is the expected behavior as the amount of available CPU is considered
when placing processes. The rejection rate depends on the request size, and how much resources are stranded. We observed
around 30\% rejections when the cluster was 75\% loaded, and each task was asking on average for 4 cores. We suspect
that the decrease in scheduling time when going from 0\% to 25\% is caused by CPU frequency scaling of the Xeon
processors \cite{intel2004speedstep}.

In our second experiment, we evaluate the scheduling time depending on the number of tasks per request. We expect that
larger requests will take more time to schedule. For this experiment, we run without any background load and gradually
increase the number of tasks per request, from 10 to 40.

Results are shown in Figure~\ref{fig:Tasks_n1}, and one can observe that the scheduling time increases with the number
of tasks per request. Although most of the Nefele control-plane code is concurrent and asynchronous, there are certain
locations where requests are handled serially. One example is the D-Bus interface between the Nefele agent and the
Process control daemon, where tasks have to be submitted for execution individually. Similar behavior has also been
reported for virtual machines when they increased the number of starting VMs~\cite{manco2017my}.

\subsection{Throughput of the scheduler}
We also evaluate how Nefele handles high throughput scenarios, where many requests arrive in a short interval. This
evaluates how well Nefele's control-plane can scale elastically, can handle high throughput, and support sub-second task
scheduling, with the goal of handling many small tasks. For these experiments, we generate a synthetic workload where
each request has on average 2 tasks that each sleep for 20 seconds.

In our third experiment requests are sent to a single Nefele node and we measure the response time, starting at a low
load that is gradually increased to 5000 requests per minute ($\approx$80 requests per second).  In the fourth
experiment, we distribute the requests over the full cluster of 15 nodes, starting at a low load that is gradually
increased to 30000 requests over a minute (500/s).

Figure~\ref{fig:n30_highThroughput} shows the results of the third experiment, when a single admission node handles the
admissions. We observe that the admission node processes the requests without queuing them, for up to 33 request per
second. The average response time in this case is less than 20 ms. Further increasing the arrival rate causes a queue to
be built up at the admission node.  Around $\approx$75 requests per second, the queue time becomes a larger fraction of
the scheduling time compared to the decision time, and therefore the response time dramatically increases to more than
300 ms at 85 requests per second.

However, as each node in Nefele is an admission node the control-plane can scale as we receive more requests by
distributing them in the cluster. Results from the fourth experiment (Figure~\ref{fig:n60_highThroughput}) shows that
when requests are distributed to multiple admission nodes throughout the cluster, Nefele can handle a significantly
larger number of requests. In this case, the 30000 requests per minute are on average handled in less than 50 ms by the
15 nodes.

\subsection{Monitoring processes}\label{sec:performance}
Finally, we evaluate the performance of the process monitoring functionality, by measuring the time from a crash in a
process to the reception of a crash notification in the monitoring process. In order to measure this value accurately
both the crashing and the monitoring process are placed on the same node, this way we do not have to synchronize
different clocks. We measure the combined detection and notification time to 4.33 ± 0.55 ms. If we immediately respawn
the crashed process upon receiving the notification, a crashed process may be restored within around 20 ms.  To put this
value in context we measured the process spawn time using other mechanisms, the results of this final experiment is
shown in Table~\ref{tab:spawning}.

On a single node, it takes around 1 ms to spawn a process using the regular shell (in this case \texttt{bash}), this
value goes up to around 17 ms when spawning with transient systemd services (using \texttt{systemd-run} from a shell).
A significant amount of time in starting a systemd service is likely spent in the setup of the D-Bus communication used
by \texttt{systemd-run} to communicate with the systemd process (as well as starting the \texttt{systemd-run} process
itself). The time to start a single process on a remote node, using Nefele, is around 15 ms, which is lower than
\texttt{systemd-run} spawn on a local node.  In both cases, D-Bus is used to communicate with the systemd process, in
the Nefele case through the control-plane on a remote node. However, in this case, a D-Bus connection has already been
established which could account for the lower value. Finally, we measure the time to create a container locally and run
a process inside using \texttt{docker run}. As expected, this is an order of magnitude slower, as filesystems have to be
mounted, namespaces created, etc. In conclusion, detecting crashes in individual processes and starting a replacement
process can be significantly faster than starting whole containers in response to failures.
\begin{table}
  \caption{Time to start a process, in milliseconds}\label{tab:spawning}
  \begin{center}
    \begin{tabular}{lr}
      \toprule
      Shell (\texttt{bash}) & $1.29 \pm 0.59$ ms \\
      \texttt{systemd-run} &  $17.3 \pm 1.43$ ms \\
      Nefele spawn & $15.3 \pm 1.97$ ms \\
      \texttt{docker run} & $762 \pm 84.8$ ms\\
      \bottomrule
    \end{tabular}
  \end{center}
\end{table}

\section{Related work}\label{sec-relatedwork}
There are two lines of research related to our work. The first is a group of projects that provide resource management
solutions through the SSI design. The aim of the SSI model is to hide the distributed nature and heterogeneity of the
underlying machines from the user, and for this reason, resources are aggregated and presented to the user as a single
pool of resources. Depending on the level of transparency the SSI wishes to implement, different types of resources must
be aggregated and presented to the user. Examples of these resources are a single entry point, a single process space, a
single memory space, a single I/O space, and a single job management system. SSIs have been implemented both at the
kernel-level and in user-space. 

GLUnix is a user-space implementation of SSI, for a cluster of workstations~\cite{ghormely1998}. It is implemented as a
global runtime environment at the user level, leveraging available operating systems primitives as building blocks.  It
provides transparent remote execution and support for interactive parallel and sequential jobs.  However, the authors
realized that the user-mode privileges are insufficient to implement a fully transparent SSI, due to constraints around
terminal IO, signaling and device accesses. \Pontus{Add some more pros/cons here perhaps, see ghormely1998, be a bit
  more specific about implementation}

Kerrighed~\cite{Morin_2003} is an example of a kernel-level SSI operating system, designed to support parallel numerical
simulations in a multi-node setting.  The goal of Kerrighed is to provide efficient resource management, high
availability, and ease of use. Kerrighed is one of the few SSI OSes which provides cluster-wide shared memory and
thread and process migration. However, due to performance problems partly caused by the distributed shared memory, the
project never received large commercial support. Similar kernel-level SSI projects are Plan 9~\cite{pike1995plan},
OpenMosix~\cite{OpenMosix} and OpenSSI~\cite{walker2008open}.

The second group of works are efforts on resource orchestration and management, for VMs and/or containers, such as
OpenStack~\cite{openstack}, Kubernetes~\cite{kubernetes}, and Docker Swarm~\cite{docker-swarm}.  All of these frameworks
present the cluster as a single unified resource entity, with a centralized point of administration
~\cite{healy2016single}. Each of these orchestrations systems translates the users' resource requirements into actual
allocations on different nodes in the cluster and hides all the details of where to place the application. However,
an aggregated resource view is only presented to the \textit{administrators} and not to the software developer. The
developer still needs to write distributed applications and/or make use of external services to ensure the applications'
consistency and accuracy. She should also explicitly define the interactions and relationships between different
application components and different resources. 

\Pontus{We could include here metaparticle, hazelcast, ..}

\section{Conclusions and future work}\label{sec-conclusion}
In this paper, we presented Nefele, a decentralized process orchestration system, inspired by SSI and Erlang/OTP.\@ The
aim of Nefele is to simplify building cloud-native applications by providing aggregated views of underlying distributed
resources as well as of dynamic run-time information, e.g., an aggregated list of running processes.
Using Nefele, the developer can programmatically deploy and manage an application on a cluster, partly
abstracted as a single node. This is achieved through a distributed control-plane, which coordinates and communicates
with an application's processes across several nodes. Nefele performs distributed process management, allowing the
developer to, e.g.,\ spawn, list, monitor failures, signal, and control processes in a distributed environment.

In Nefele, relationships between different processes are defined in the chosen programming language, rather than
externally through, e.g., a YAML template.  The programmatic way of defining relationships makes it possible to
customize the logic behind them, e.g., the developer can implement custom logic for determining when and how to scale.
 
Our evaluations show that Nefele can deploy, scale, and manage distributed processes effectively, with an average
process scheduling time between 10 and 20 ms, depending on task size and arrival rate. In a cluster of 15 nodes, Nefele
is able to handle 30000 requests per minute with an average scheduling time below 50 ms.

Since Nefele's control-plane is decentralized, there is no single point of failure and the system can remain operational
through failures such as network partitioning. The process fault-tolerance model using supervision trees, works well
with a failure detection and handling time as low as 20 ms.

Currently, an alpha release of Nefele is running as a service in our datacenter, open for internal users.  Our users are
deploying multi-tiered distributed applications, testing the development- and deployment advantages of Nefele, and
specifically taking advantage of the built-in IPC and failover mechanisms. We are extending Nefele with more features,
adding more internal and external services that further simplifies the construction of distributed applications.  We are
also testing the system in larger, more heterogeneous clusters, in order to discover and fix bottle-necks. Based on user
feedback, we plan to further refine the APIs and improve the performance of existing functionality, e.g., through
applying machine intelligence in the Placement agent and reduce IPC latency through~RDMA.

\bibliographystyle{IEEEtran}
\bibliography{nefele-paper}

\newpage

\begin{IEEEbiography}[{\includegraphics[width=1in,height=1.25in,clip,keepaspectratio]{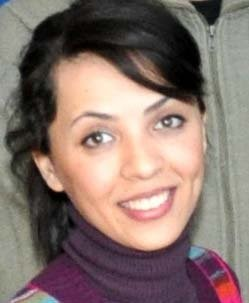}}]%
{Mina Sedaghat}
is an experienced researcher at Ericsson Research, Stockholm. Her main interest lies in distributed systems, cloud and Edge computing, and specifically building self-managed distributed systems. She has a PhD in Computer Science from Umea university, Sweden, and she is the author of several scientific papers.
\end{IEEEbiography}
\vskip 0pt plus -1fil
\begin{IEEEbiography}[{\includegraphics[width=1in,height=1.25in,clip,keepaspectratio]{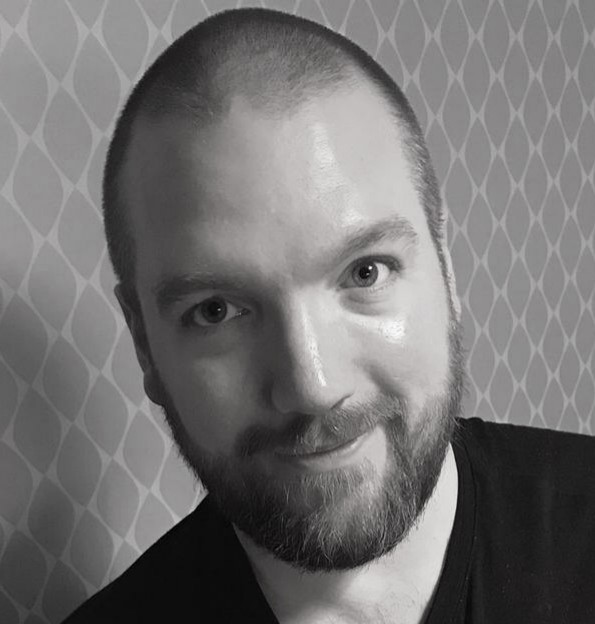}}]%
{Pontus Sköldström}
(MsC Communication Systems, 2008) works at Ericsson Research as a Senior Researcher. His current research is on distributed systems, with a focus on software solutions for distributed Cloud and Edge computing, while in the past he worked mainly on optical- and packet networks and their integration. 
\end{IEEEbiography}
\vskip 0pt plus -1fil
\begin{IEEEbiography}[{\includegraphics[width=1in,height=1.25in,clip,keepaspectratio]{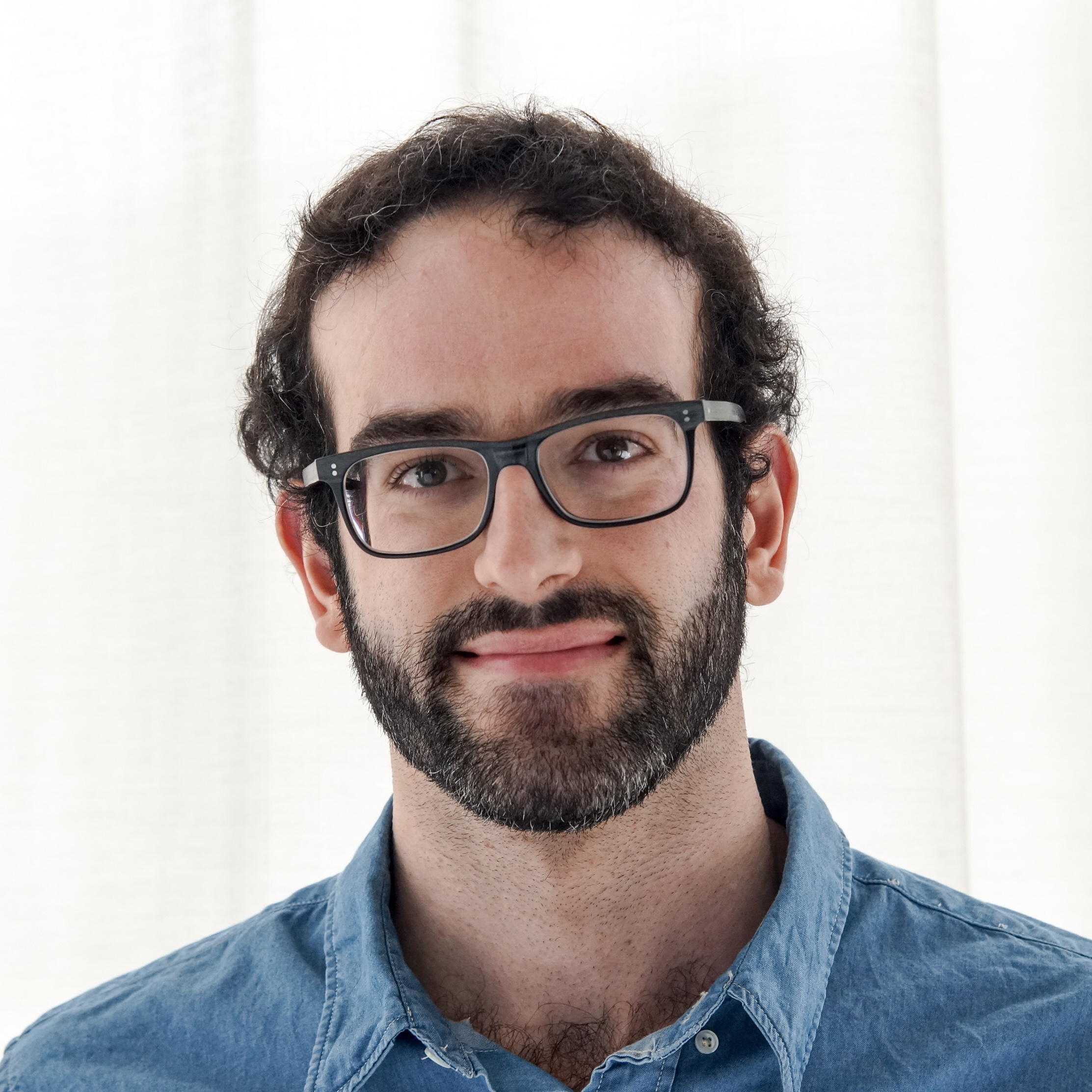}}]%
{Daniel Turull}
 works as Senior Researcher in Ericsson Research since 2014. He holds a M.Sc. degree in Telecommunications engineering from the Universitat Politècnica de Catalunya, Spain, in 2010, and a Licentiate degree in communication systems from the Royal Institute of Technology, Sweden, in 2016. He has worked with the Linux kernel, OpenFlow, virtualization technologies as well as disaggregated cloud infrastructure. He has co-authored several scientific papers and patents.
\end{IEEEbiography}
\vskip 0pt plus -1fil
\begin{IEEEbiography}[{\includegraphics[width=1in,height=1.25in,clip,keepaspectratio]{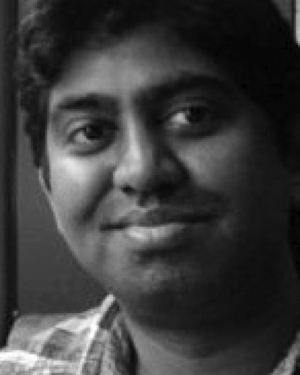}}]%
{Vinay Yadhav}
 received the M.Sc. degree in communication systems from the Royal Institute of Technology, Sweden. He joined Ericsson in 2012. He is a Senior Researcher of cloud technologies with Ericsson Research. He has researched on projects ranging from Networking-as-a-Service in cloud platforms, federated cloud research and distributed service deployments across heterogeneous cloud platforms, service migration in mobile networks to memory disaggregation. He has contributed to OpenStack under the Tap-as-a-Service project and has co-authored several patent applications.
\end{IEEEbiography}
\vskip 0pt plus -1fil
\begin{IEEEbiography}[{\includegraphics[width=1in,height=1.25in,clip,keepaspectratio]{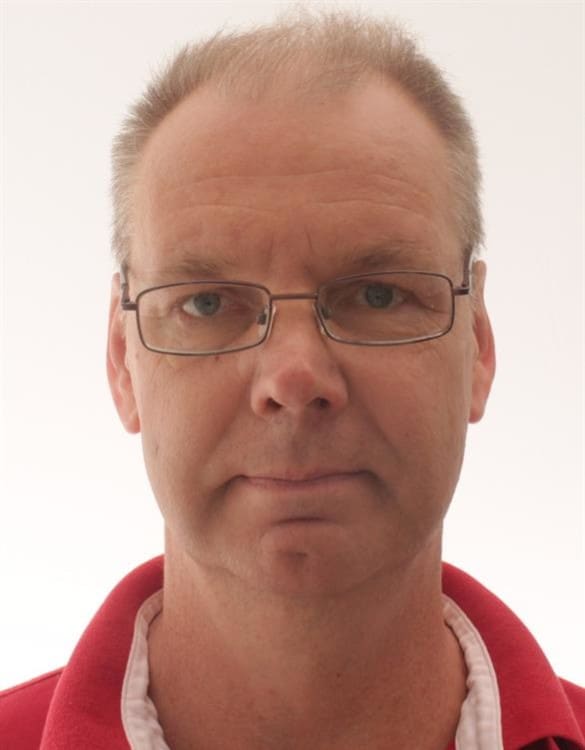}}]%
{Joacim Halén}
 works at Ericsson Research as an expert researcher. His work focuses on distributed software design for cloud systems. He has co-authored several scientific papers as well as several patents.
\end{IEEEbiography}
\vskip 0pt plus -2fil
\begin{IEEEbiography}[{\includegraphics[width=1.1in,height=1.25in,clip,keepaspectratio]{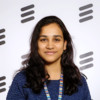}}]%
{Madhubala Ganesan}
 is a researcher at Ericsson Research, Stockholm. She received her M.Sc  from Leeds Beckett University. She is interested in Distributed Systems, Cloud Computing and IoT.
\end{IEEEbiography}
\vskip 0pt plus -1fil
\begin{IEEEbiography}[{\includegraphics[width=1in,height=1.25in,clip,keepaspectratio]{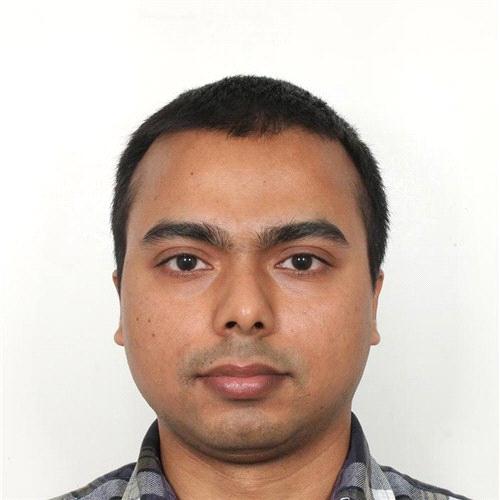}}]%
{Amardeep Mehta}
works as a researcher at Ericsson Research, Stockholm. He received his M.Sc. in Computer Science from Uppsala University and Ph.D. in Computing Science from Umeå University, Sweden. His research interests include workload analysis and resource management problems for distributed systems.
\end{IEEEbiography}
\vskip 0pt plus -2fil
\begin{IEEEbiography}[{\includegraphics[width=1in,height=1.25in,clip,keepaspectratio]{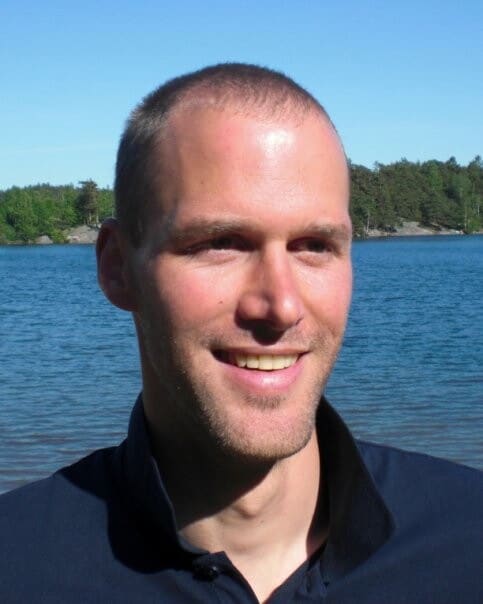}}]%
{Wolfgang John}
 is a Research Leader at Ericsson Research in Kista, Stockholm. His current research focus lies primarily on distributed Cloud and Edge computing system concepts for both telco and IT applications. Wolfgang holds a PhD (2010) in Computer Engineering from Chalmers University of Technology, Gothenburg, Sweden, and he has co-authored over 50 scientific papers and reports as well as several patents.
\end{IEEEbiography}

\end{document}